\documentclass[prl,aps,twocolumn,footinbib,showpacs,preprintnumbers,superscriptaddress]{revtex4-1}

\usepackage{graphicx}
\usepackage{dcolumn}
\usepackage{bm}
\usepackage{amssymb}
\usepackage{pifont}
\usepackage{amsmath}
\usepackage{times}
\usepackage[nooneline]{subfigure}

\begin{document}

\title{Shear diversity prevents collective synchronization}
\author{Ernest Montbri\'o}
\affiliation{Department of Information and Communication Technologies,
Universitat Pompeu Fabra, 08018 Barcelona, Spain}
\author{Diego Paz\'o}
\affiliation{Instituto de F\'{i}sica de Cantabria (IFCA), CSIC-Universidad de
Cantabria, 39005 Santander, Spain}

\date{\today}

\begin{abstract}
Large ensembles of heterogeneous oscillators often 
exhibit collective synchronization as a result of mutual
interactions. If the oscillators have distributed natural 
frequencies and common shear (or nonisochronicity), 
the transition from incoherence to collective synchronization 
is known to occur at large enough values of the coupling strength.
However, here we demonstrate that shear diversity 
cannot be counterbalanced by diffusive coupling leading to  synchronization. 
We present the first analytical results for the Kuramoto model with distributed shear, 
and show that the onset of collective synchronization is impossible if 
the width of the shear distribution exceeds a precise threshold.

\end{abstract}
\maketitle 

Collective synchronization is a form of self-organization in time that results 
from  the interactions among a large heterogeneous
population of self-sustained oscillators
\cite{Win67,Kur84,PRK01,MMZ04,ABP+05}. This phenomenon is 
observed in a large variety of systems that range
from biology to chemistry, 
physics and engineering (see e.g.~\cite{Str03}).
For the sake of mathematical simplicity,
most theoretical advances in 
this field consider oscillators with different natural frequencies.
Nevertheless, it is of great importance 
to know how other sources of heterogeneity  
influence collective synchronization.
This question has been addressed 
in the cases of heterogeneous patterns of connectivity \cite{coupling} 
and interaction delays \cite{delay}.

The so-called shear (or nonisochronicity)  is a crucial 
nonlinear ingredient for the formation of patterns 
in oscillatory extended media \cite{Kur84,CH93}, 
as well as for the onset of complex behaviors in ensembles  
of identical limit-cycle oscillators \cite{HR92}.
Although shear is a generic feature of  oscillators, 
studies considering shear diversity in ensembles of oscillators
are very scarce and focus on a regime far from synchronization \cite{BMK03}. 
Here we will show that distributed shear plays a
key role in collective synchronization
and may even prevent its onset.

Mathematical formulations of collective synchronization
usually consider as elementary oscillatory unit 
a normal form describing a
system  near the onset of
oscillations (via the Hopf bifurcation), 
the so-called Stuart-Landau (SL) oscillator \cite{Kur84}:
\begin{equation}
\dot \varrho  =   \varrho(1-\varrho^2),\quad 
\dot \theta  = \omega +q (1-\varrho^2).
\label{SL}
\end{equation}
Here the natural frequency $\omega$ determines the 
frequency of rotation on the attractor of radius $\varrho(t)=1$. 
Additionally $q$ quantifies the shear of the flow,
i.e.~how much perturbations off the limit cycle modify the angular 
frequency $\dot \theta$.  It is then 
usual to consider an ensemble of $N \gg 1$ globally 
coupled SL oscillators \eqref{SL},
a mean-field version of the complex Ginzburg-Landau equation with disorder 
\cite{Kur75}:
\begin{equation}
\dot{z}_j = z_j \left[1 + i(\omega_j + q_j) - ( 1 + i q_j) |z_j|^2\right]
+ K(1+i c_1) (Z- z_j) ,
\label{Complex2}
\end{equation}
where $z_j= \varrho_j e^{i\theta_j}$, $j=1,\ldots,N$,
and $Z=N^{-1}\sum_{k=1}^{N} z_k$. 

Previous works studying model \eqref{Complex2} have adopted the simplifying assumption 
that diversity is only present in the natural 
frequencies $\omega_j$,  and 
the shear is either absent $q_j=0$ 
\cite{Kur75,aizawa76}, or constant $q_j=q$
\cite{cross}.
However, in a heterogeneous ensemble, either the inherent 
(say genetic variability for living organisms, or 
tolerances for electronic circuits)
or the imposed (e.g.~experiments using coupled chemical 
reactors \cite{chem}) disorder 
will generically be reflected in both natural frequency 
and shear terms.

The aim of this Letter is to analyze the genuine problem of 
collective synchronization in a large ensemble of SL 
oscillators~\eqref{Complex2} with $\omega_j$ and $q_j$ distributed.
In our mathematical analysis we will assume that the oscillators are 
weakly coupled, i.e. $|K|$ is small. In such case, the dynamics of 
Eq.~\eqref{Complex2} is well described by the phases only \cite{Kur84},
\begin{equation}
\dot{\theta_j}  = \omega_j + K  q_j   + K R [ \sin(\Psi-\theta_j)  - 
q_j \cos(\Psi-\theta_j)  ].
\label{model}
\end{equation}
Here, the complex order parameter $r =R e^{i \Psi} = N^{-1} \sum_{k=1}^N e^{i \theta_k}$
is a mean field and 
measures the degree of synchronization in the population. 
For the sake of simplicity, we assume 
a purely dissipative coupling, $c_1=0$.
In Eq.~\eqref{model}, the well-known 
Kuramoto model is recovered in the fully isochronous case, 
$q_j=0$ \cite{Kur75,Kur84,Str00,ABP+05}, 
whereas 
the nonisochronous case without disorder, $q_j=q$, 
corresponds to the so-called Sakaguchi-Kuramoto model~\cite{SK86}
\footnote{Defining $\tan \beta_j=q_j$ and $|\beta_j|\le\tfrac{\pi}{2}$, 
Eq.~(\ref{model}) reads $\dot{\theta_j}  =  \omega_j + 
K  \tan \beta_j + \tfrac{K}{\cos \beta_j} R \sin(\Psi-\theta_j-\beta_j)$, 
and one can recognize the equivalence with the Sakaguchi-Kuramoto model when $\beta_j=\beta$.}.

To analyze model \eqref{model} we adopt its thermodynamic limit
$N\rightarrow \infty$. Thus we drop the indices 
and introduce the probability density for the phases
$f(\theta,\omega,q,t)$ \cite{SM91}. 
Then, the quantity $f(\theta,\omega,q,t)\, d\theta \, d\omega \, dq$
represents the ratio of oscillators with phases between 
$\theta$ and $\theta+d\theta$,
natural frequencies between $\omega$ and $\omega+d\omega$, and shear
between $q$ and $q+dq$.
The density $f$ obeys the continuity equation 
\begin{equation}
\partial_t f= - \partial_\theta
\left( \left\{\omega+K q +  \frac{ K}{2i}\left[r e^{-i \theta}(1 - i q) - 
{\rm c.c.}\right] \right\} f \right),
\label{cont_eq}
\end{equation}
where c.c.~stands for complex conjugate of the preceding term, and the complex order parameter is
\begin{equation}
r=\int_{-\infty}^{\infty} \int_{-\infty}^{\infty} \int_0^{2\pi} e^{i \theta}
f(\theta,\omega,q,t)~d\theta ~d\omega~  dq,
\label{z}
\end{equation}
If the phases
are uniformly distributed $r$ vanishes. This state is
customarily referred to as incoherence. 
Since $f(\theta,\omega,q,t)$ is real and $2\pi$-periodic
in the $\theta$ variable, it admits the Fourier expansion
\begin{equation}
f(\theta,\omega,q,t)=\frac{p(\omega,q)}{2\pi}
\sum_{l=-\infty}^\infty  f_l(\omega,q,t) e^{il\theta} 
\label{fourier}
\end{equation}
where $f_l=f_{-l}^*$, $f_{0}=1$, 
and $p(\omega,q)$ is the joint probability density function (pdf) of 
$\omega$ and $q$. The first Fourier mode is important because it
determines the order parameter \eqref{z}:
\begin{equation}
r^*(t)=\iint_{-\infty}^{\infty}  p(\omega,q) f_1
(\omega,q,t) ~ d\omega ~ dq .
\label{z1}
\end{equation}
Inserting the Fourier series (\ref{fourier}) into the continuity 
Eq.~(\ref{cont_eq}) we obtain the
following set of integro-differential equations for the
Fourier modes
\begin{equation}
\partial_t 
f_l =-i l (\omega+Kq)  f_l  
+\frac{K l}{2} \left[ r^* (1+i q) f_{l-1} - r (1-i q) f_{l+1}  \right] .
\label{fourier_set}
\end{equation}
Recently Ott and Antonsen (OA) found that the ansatz  \cite{OA08,OA09}
\begin{equation}
f_l(\omega,q,t)=\alpha(\omega,q,t)^l,
\label{ansatz}
\end{equation}
is a particular solution of the Kuramoto model and 
related systems with distributed natural 
frequencies. Here we also resort to \eqref{ansatz} as it 
turns out to be a solution in our case if $\alpha$ obeys
\begin{equation}
\partial_t \alpha  = -i(\omega+ K q) \alpha + \frac{K}{2} \left[  r^* (1+i q)   
 - r \left(1-i q\right) \alpha^{2} \right] 
\label{alpha} 
\end{equation}

The idea behind Ott and Antonsen's approach is
to simplify an infinite set of equations---like Eq.~\eqref{alpha}---using
distributions that can be inserted in 
Eq.~\eqref{z1} and integrated via Cauchy's residue theorem (see below).

In this Letter we assume that $\omega$ and $q$ are independent
random variables, $p(\omega,q)=g(\omega) h(q)$. 
Moreover we restrict the analysis to
symmetric unimodal pdfs $g(\omega)$ and $h(q)$
centered at $\omega_0$ and $q_0$, respectively.
We start choosing $g$ and $h$ to be Lorentzian (Cauchy) pdfs, 
\begin{equation}
g(\omega)= \frac{\delta/\pi}{(\omega-\omega_0)^2+\delta^2};~~h(q)=
\frac{\gamma/\pi}{(q-q_0)^2+\gamma^2}.
\label{lorentzians}
\end{equation}
The integrals in \eqref{z1} can be evaluated by means of the residue theorem
with the contour closings at infinity in the lower
or the upper half plane of $\mathbb{C}$,
granted $\alpha=f_1$ can be continued from real $\omega$ and $q$
into complex $\omega=\omega_r+i\omega_i$ and $q=q_r+iq_i$.

Regarding variable $\omega$, analyticity of $\alpha$ holds in the
lower half complex $\omega$-plane (see \cite{OA08}).
As $g(\omega)=(2\pi i)^{-1} [(\omega-\omega_0-i\delta)^{-1}-
(\omega-\omega_0+i\delta)^{-1}]$
has only one simple pole $\omega^p=\omega_0-i\delta$ inside this integration contour,
only the value of $\alpha=f_1$ at $\omega=\omega^p$ 
counts in the integral over $\omega$ in Eq.~\eqref{z1}.

The integration over $q$ in Eq.~\eqref{z1} is more intricate.
We have to choose an integration contour such 
that, if  $\alpha$ is analytic
and $|\alpha|\le 1$ everywhere inside the contour at $t=0$, 
this will hold for all $t>0$.
As $\alpha$ is a solution of Eq.~\eqref{alpha}, the analyticity of $\alpha$ at $t=0$
is  preserved as $t$ grows if $\alpha$ remains finite \cite{CL}.
Moreover, if $\alpha$ is analytic, the Cauchy-Riemann conditions
imply $\partial_{q_r} |\alpha|+\partial_{q_i}|\alpha|\ge 0$,
and the maximum of $|\alpha|$ is necessarily located on the boundary
(namely, on the integration contour).
First of all, setting $\alpha=|\alpha|e^{-i\psi}$
in Eq.~\eqref{alpha}, we obtain that, on the real $q$-axis,
$|\alpha|$ is governed by
\begin{equation}
\partial_t |\alpha| = -\delta |\alpha|+\frac{K}{2} \, \mbox{Re}\left[r^* e^{i\psi} 
(1+i q)\right]\left( 1 -|\alpha|^2 \right) .
\label{abs_alpha}
\end{equation}
The fact that $\partial_t |\alpha|=-\delta<0$
at $|\alpha|=1$ guarantees that,
if an initial condition satisfies $|\alpha(\omega,q,t=0)|<1$,
this will hold for all $t>0$.
Consequently the series in Eq.~\eqref{fourier} remains convergent.
Regarding the line integral along the semicircular path
$q=|q|e^{i\vartheta}$ with $|q|\to\infty$, Eq.~\eqref{alpha} yields 
\begin{equation}
\frac{\partial_t |\alpha|}{K|q|} = |\alpha| \sin\vartheta 
-\frac{R}{2}\left[ \sin(\vartheta+\chi) + |\alpha|^2 \sin(\vartheta -\chi)\right] 
\label{eqinf}
\end{equation}
where $\chi(\omega,q,t)=\psi(\omega,q,t)-\Psi(t)$. 
At $|\alpha|=1$ Eq.~\eqref{eqinf} becomes 
\begin{equation}
\partial_t |\alpha| =  (1 - R \,\cos \chi )  K |q| \sin \vartheta
\label{eqsimple} 
\end{equation}
In this equation the desired relation, $\partial_t |\alpha| \le 0$, 
is fulfilled 
in the lower half complex $q$-plane $\vartheta \in (-\pi,0)$
only  if $K>0$, and in the upper half-plane
$\vartheta \in (0,\pi)$ for $K<0$.

The integral over $q$ in~\eqref{z1} can be now conveniently evaluated, 
and yields 
\begin{equation}
r^*(t)=\alpha(\omega=\omega^p,q=q^p,t) = a(t)
\end{equation}
with $q^p=q_0- i\gamma$ for $K>0$, and $q^p=q_0 + i \gamma$ for $K<0$.
Thus, among the infinite set of equations, Eq.~\eqref{alpha},
only the one at $(\omega,q)=(\omega^p,q^p)$ is needed:
\begin{equation}
\dot a= -i\omega^p a+ \frac{K}{2} (1 -iq^p)(1-|a|^2) a.
\label{alpha2}
\end{equation}
The dynamics of the radial component $|a|= R$ obeys 
\begin{eqnarray}
\dot R &=& \left[ -\delta + \frac{K}{2}(1\mp\gamma)(1-R^2) \right] R \label{rho_eq}
\end{eqnarray}
where the ``$\mp$'' stands for ``$-$'' if $K>0$, and ``$+$'' if $K<0$.
The incoherent state, $R=0$, is always stable except above the line
\begin{equation}
K_c= \frac{2\delta}{1-\gamma} \, \text{  if $\gamma < 1$ }.
\label{Kc0}
\end{equation}
At $K_c$ a stable nontrivial solution, corresponding to a partially synchronized state, 
appears with
\begin{equation}
R^2= \frac{K-K_c}{K}
\label{rhos}
\end{equation}
Equations~\eqref{rhos} and \eqref{Kc0} are depicted in Fig.~1(a) 
and compared with numerical simulations of Eq.~\eqref{Complex2}, i.e.~using SL oscillators 
\footnote{We have also performed numerical simulations using phase oscillators, 
Eq.~(\ref{model}). There we find a perfect agreement with 
our analytical results.}.
In the Kuramoto model---recovered when $\gamma=0$---a large enough coupling strength $K$
\emph{always} results in partial synchronization of the population above $K_c$,
for any width $\delta$ of the frequency distribution 
$g(\omega)$. However, here we find that the width $\gamma$ 
of the shear distribution $h(q)$ has a more severe effect 
on the synchronization transition. If $\gamma \geq \gamma_d= 1$, 
synchrony disappears for all $K$ and $\delta$, 
and incoherence  becomes the only stable state.
It is noteworthy that this is a collective phenomenon caused 
by the presence of distributed shear, and it  has no counterpart 
in the case of two coupled SL oscillators 
\footnote{The phase reduction of Eq.~(\ref{Complex2}) for $N=2$ is not 
Eq.~(\ref{model}), see \cite{Kur84}.} \cite{AEK90,BMK03}.

\begin{figure}
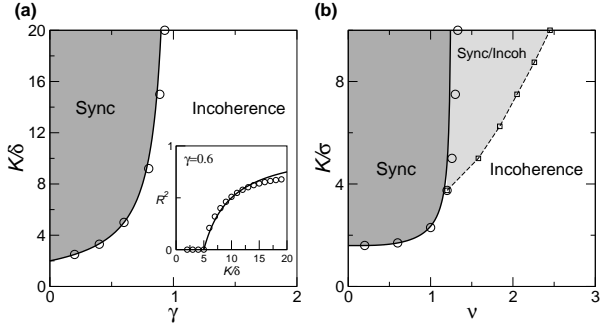

\centerline{\includegraphics *[width=79mm,clip=true]{Fig1.eps}}
\caption{Phase diagram for
(a) Lorentzian pdfs
$g(\omega)$ and $h(q)$, Eq.~\eqref{lorentzians},
and (b) Gaussian pdfs.
(a) Solid line: synchronization critical coupling,
Eq.~\eqref{Kc0}. Inset: Order parameter as a function of $K$
for $\gamma=0.6$; the solid line obeys Eq.~\eqref{rhos}. 
(b) Solid line: critical coupling given by Eq.~\eqref{Kgaussian}. 
In both panels, (a) and (b),
the symbols correspond to numerical results obtained using an
ensemble of SL oscillators, Eq.~\eqref{Complex2}, with parameters: (a) $N=40000$,  
$\delta=0.01, \omega_0=q_0=\tfrac{1}{2}$; (b) $N=22500$, $\sigma=0.02$, 
$\omega_0=q_0=0$. Parameters $\omega_j$ and $q_j$ were 
deterministically selected to represent
$p(\omega,q)$ and averages were done over time.}
\end{figure}

To confirm the generality of these findings for other pdfs \cite{LCT10},
next we follow \cite{SM91} and perform the 
linear stability analysis of Eqs.~\eqref{fourier_set}
about the incoherent state $f_{l\ne 0}=0$. 
We find that the only potentially unstable modes are $l=\pm 1$.
Inserting $f_1(\omega,q,t)= b(\omega,q)e^{\lambda t}$ into Eq.~\eqref{fourier_set},
the discrete spectrum of eigenvalues $\lambda$ can
be obtained by virtue of a self-consistency argument. 
This yields the integral equation   
\begin{equation}
\frac{2}{K}=   \iint_{-\infty}^{\infty} 
\frac{1+iq}{\lambda+i(\omega+q K)} g(\omega) h(q)~ d\omega~dq.
\label{incoherence}
\end{equation}
The border of unstable incoherence $K_c$, is found imposing 
$\mbox{Re}(\lambda)\to 0^+$. If $g(\omega)$ and $h(q)$ are Gaussian 
functions with variances $\sigma^2$ and $\nu^2$, respectively,
and $h(q)$ has zero mean ($q_0=0$), the critical coupling can be explicitly obtained:
\begin{equation}
K_c= \sigma \sqrt{ \frac{-\pi+8 \nu^2 + \sqrt{\pi(\pi+16\nu^2)}}{2\nu^2(\pi-2\nu^2)}}. 
\label{Kgaussian}
\end{equation}
This function is plotted in Fig.~1(b) and compared with the results 
of numerical simulations of 
Eq.~\eqref{Complex2}.
Remarkably, we find again a threshold for the dispersion 
of $h(q)$, $\nu_d=\sqrt{\pi/2}=1.253\ldots$, above which 
incoherence is stable for all $K$. 

We have found that the divergence of $K_c$
does not only exist  
for Lorentzian and Gaussian distributions, but for 
any symmetric unimodal distribution.
This divergence occurs at a shear diversity that is conveniently expressed in 
terms 
of the peak value $h(q_0)$. If $q_0=0$, the divergence occurs at
\begin{equation}
h(0)= \frac{1}{\pi}. 
\label{divergence}
\end{equation}
Otherwise, if $h(q)$ is not centered at zero,
$K_c$ also diverges at a certain value of $h(q_0)=h_d$,
but a simple distribution-independent formula like \eqref{divergence} 
does not exist \footnote{For most distributions we have investigated, by means 
of
Eq.~(\ref{incoherence}), $h_d$ decreases as
$|q_0|$ grows. Only for the Lorentzian 
distribution $h_d=\pi^{-1}$ holds for all $q_0$ values.}. 

Our numerical calculations of Eq.~\eqref{Complex2} using Gaussian pdfs 
also reveal that incoherence and synchronization coexist in a region with 
large $K/\sigma$ [light-shaded region, Fig.~1(b)]. 
Note that this region is not present  
in the Lorentzian case [Fig.~1(a)], in spite of the similar 
bell-shaped form of these two distributions.
Figure~1(b)
suggests that the destabilization of incoherence
may also occur through a subcritical bifurcation for certain pdfs. 
To elucidate the supercritical or subcritical character of the
synchronization transition,
we carry out a self-consistency analysis 
``{\em \`a la} Kuramoto'' \cite{Kur75,Kur84,SK86,Str00} in the limit
of large coupling and/or very small frequency dispersion, i.e.~$K g(\omega_0) 
\gg1$. After going into a rotating framework
$\theta\to\theta+\omega_0 t$, rescaling time $t\to K^{-1} t$, and 
neglecting the $\omega_j/K$ term, we
approximate Eq.~\eqref{model} by:
\begin{equation}
 \dot \theta_j= q_j + R [\sin(\Psi-\theta_j)-q_j \cos(\Psi-\theta_j)].
\label{1}
\end{equation}
Hereafter we assume $q_0=0$.
In a partially synchronized state
the population splits into two groups, the synchronized
(or locked) subpopulation with $|q|\le q_{\rm max}=R/\sqrt{1-R^2}$, 
and the desynchronized (or drifting)
one with $|q|>q_{\rm max}$.
Both of subpopulations do contribute to the order parameter:
\begin{equation}
R =\left< \cos \theta\right>_s + \left< \cos\theta\right>_{ds}
\label{sc}
\end{equation}
where we have chosen a reference frame where $\Psi=0$
and the brackets denote averages over each subpopulation.
We can now make an expansion in powers of $R$ for each contribution.
Up to cubic order we obtain:
\begin{eqnarray}
\left< \cos \theta \right>_{s} \simeq
R\frac{\pi}{2} h(0) + R^2 \frac{2}{3}h(0)  + R^3\frac{\pi}{8}\left[h(0) + 
\frac{h''(0)}{2}  \right]  & & \nonumber\\
\left<\cos\theta\right>_{ds} \simeq \frac{R}{2} - R^2 \frac{2}{3} h(0) + 
\frac{R^3}{4}\left[\frac{1}{2}+\int_0^\infty \frac{h'(q)}{q} \, 
dq  \right] & & \nonumber
\label{o3}
\end{eqnarray}
where we have assumed $h(q)$ is twice differentiable at the origin.
Inserting these expansions into Eq.~\eqref{sc}, and
discarding the trivial solution $R=0$ 
we find that $R$ follows asymptotically a square-root dependence of the form:
\begin{equation}
 R=\sqrt{\frac{1-\pi h(0)}{J}}
\label{rj}
\end{equation}
where
\begin{equation}
J=\frac{1}{2}\left[1+\frac{\pi h''(0)}{4} +\int_0^\infty \frac{h'(q)}{q} \, dq \right]
\label{j}
 \end{equation}
is evaluated at $h(0)=\pi^{-1}$.
The sign of $J$ determines the orientation of the bifurcating branch
in Eq.~\eqref{rj}, as showed in the conjectured bifurcation 
scenarios in Fig.~2(a,b). When $J>0$, 
as for the Gaussian distribution ($J=\tfrac{1}{2}-\tfrac{3}{4\pi}=0.261\ldots$), 
a partially synchronized solution branches 
off from incoherence subcritically.
This is in concordance with the numerical 
results in Fig.~1(b). The scenario for $J<0$, Fig.~2(b),
is also followed by pdfs with a
non-differentiable maximum of type $h(q)=h(0) + h'(0^+) |q|+\cdots$, 
like the triangular or Laplace distributions (formally $J=-\infty$). 
It is important to note that 
bistability incoherence-synchronization is found
irrespective of the sign of $J$,
because Eq.~\eqref{1} has a stable fixed point 
at $\theta_j=\Psi$ ($R=1$) that persists 
in the form of a solution with $R$ near $1$, 
provided $g(\omega)$ has a small dispersion (or $K$ is large enough).

The case of Lorentzian $h(q)$ is quite peculiar. 
On the one hand, $J$ vanishes for this pdf, what is consistent
with the infinitely abrupt transition 
predicted by the OA ansatz in the limit $K\to\infty$, see Fig.~2(c).  
On the other hand, according to our numerical simulations, 
the synchronized solution $R=1$ of Eq.~\eqref{1}
does not persist for $\gamma>1$ if
$g(\omega)$ is not a delta function.
This indicates that, for heavy-tailed $h(q)$, 
the term $\omega_j/K$ neglected in 
Eq.~\eqref{1} may become relevant and destroy the synchronized 
solution beyond a certain critical value of $h(q_0)^{-1}$.

\begin{figure}
\centerline{\includegraphics *[width=80mm,clip=true]{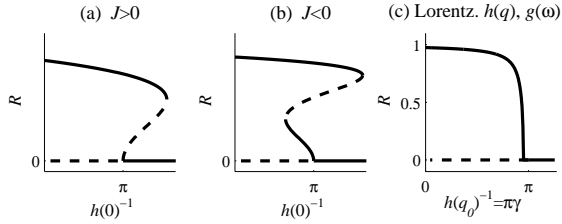}}
\caption{(a,b) Sketch of the possible 
bifurcation scenarios 
for large $K$ values, depending on the sign of $J$, Eq.~\eqref{j}.
Solid (dashed) lines indicate stable (unstable) solutions.
(c) Bifurcation diagram for Lorentzian pdfs at $K/\delta=40$, obtained 
from Eqs.~\eqref{rhos} and \eqref{Kc0}.}
\end{figure}

In summary, this Letter
uncovers the effect of shear diversity 
on the collective synchronization of globally coupled oscillators.  
We have obtained the first analytical results for the Kuramoto model 
with distributed shear \eqref{model}. They show that, if 
shear is widely distributed,  
incoherence is always stable and for some distributions
---such as the Lorentzian one---
synchronization is impossible. The techniques used here 
can be readily applied to a number of extensions of the
model \eqref{model}, 
such as considering other distributions $p(\omega,q)$, periodic 
or stochastic driving, time delays, 
or networks and communities of oscillator populations.

\begin{acknowledgments}
We thank A. Ledberg, J.~M. L\'opez and M.~A. Mat\'ias for their 
critical reading of our manuscript.
DP acknowledges support by CSIC under the JAE-Doc Programme, and
from the MICINN project No.~FIS2009-12964-C05-05.
\end{acknowledgments}

\bibliographystyle{prsty}

\end{document}